# Experimental neck muscle pain impairs standing balance in humans

Nicolas VUILLERME and Nicolas PINSAULT

Address for correspondence:

Nicolas VUILLERME

Laboratoire TIMC-IMAG, UMR UJF CNRS 5525,

Faculté de Médecine,

38706 La Tronche cédex, France.

Fax: (33) (0) 4 76 51 86 67

Email address: nicolas.vuillerme@imag.fr

**Abstract**

Impaired postural control has been reported in patients with chronic neck pain of both traumatic and non-traumatic aetiologies, but whether neck muscle painful stimulation *per se* can affect balance control during quiet standing in humans remains unclear. The purpose of the present experiment was thus to investigate the effect of experimental neck muscle pain on standing balance in young healthy adults. To achieve this goal, sixteen male university students were asked to stand upright as still as possible on a force platform with their eyes closed in two conditions of No pain and Pain of the neck muscles elicited by experimental painful electrical stimulation. Postural control and postural performance were assessed by the displacements of the centre of foot pressure (CoP) and of the centre of mass (CoM), respectively. Results showed increased CoP and CoM displacements variance, range, mean velocity, and mean and median frequencies in the Pain relative to the No pain condition. The present findings emphasize the destabilizing effect of experimental neck muscle pain *per se* and, more largely, stress the importance of intact neck neuromuscular function on standing balance.

**Introduction**

The ability to maintain an upright stance is known to be critical for both the acquisition and control of motor abilities and represents an essential requirement for physical and daily activities. Postural control is a complex task that requires the integration of visual, vestibular and somatosensory inputs from all over the body to assess the position and motion of the body in space and the ability to generate forces to control body position (e.g., Massion 1994). When any of these components is altered either by aging, pathology, injury or by experimental manipulations in healthy individuals, postural control generally decreases. For instance, impaired postural control during quiet standing has been reported in patients with chronic neck pain of different aetiologies (whiplash-related and insidious) (e.g., Alund et al., 1993; Dehner et al., 2008; Karlberg et al., 1995; Madeleine et al., 2004; Michaelson et al., 2003; Persson et al., 1996; Poole et al., 2008; Treleaven et al., 2005). Although the exact mechanisms are difficult to answer, the observed postural impairments control has been suggested to stem from disturbances of the neck proprioceptive input to the central nervous system and/or the central processing of such input associated with chronic neck pain (Brandt and Bronstein, 2001; Michaelson et al., 2003; Persson et al., 1996). In accordance with this assumption, numerous investigations have documented deficits in neck proprioceptive acuity in patients suffering from chronic neck pain of both traumatic and non-traumatic aetiologies (e.g., Heikkila and Astrom, 1996; Kristjansson et al., 2003; Pnisault et al., 2008; Revel et al., 1991; Treleaven et al., 2003). At this point, however, as recently mentioned by Bennell and co-workers (Bennell and Hinman, 2005; Bennell et al., 2005), the extent to which postural impairments reported in symptomatic patients are due to the effect of painful stimulation itself and/or to other features associated with the pathology is difficult to establish. Indeed, depending on the pathology but also age of the subjects, specific sensory, motor and/or cognitive disorders and/or adaptive adjustments of motor control strategies might emerge (e.g., Latash and Anson, 1996; Vuillerme et al., 2001), thus inducing a modification of postural behaviour. Within this context, the purpose of the present experiment was to determine whether neck muscle pain *per se* affects standing in humans. To isolate systematically the influence of muscle pain, an experimental muscle pain model (Madeleine et al., 1999), presenting the advantage of standardizing the origin, the locus, the intensity and the duration of muscle pain (Svensson and Arendt-Nielsen, 1995), was adopted. Accordingly, we used a repeated measures within-subjects design, the neck muscle pain being experimentally induced in young healthy individuals. Furthermore, considering the dominant role played by visual information on postural control during quiet standing (e.g., see Redfern et al., 2001 for a recent review), and, specifically its effectiveness in limiting the destabilizing effect of alteration of neck neuromuscular function (e.g., Schieppati et al., 2003; Vuillerme et al., 2005), the experiment was performed in the absence of vision to better isolate the potential postural effects of experimental neck muscle pain.

**Methods**

Subjects

Sixteen male university students (mean age = 22.2 ± 1.8 years; mean body weight = 73.0 ± 11.8 kg; mean height = 181.4 ± 6.4 cm) voluntarily participated in the experiment. They gave their informed consent to the experimental procedure as required by the Helsinki declaration (1964) and the local Ethics Committee. None of the subjects presented any history of neck trauma, neck pain and any type of musculoskeletal problems, neurological disease or vestibular impairment.

Task and procedure

Subjects stood barefoot on a force platform (Equi+, model PF01, Aix les Bains, France) in a natural position (feet abducted at 30°, heels separated by 3 cm), their arms hanging loosely by their sides with their eyes closed and were asked to stand as still as possible (Zok et al., 2008). This instruction was repeated at the beginning of every trial.

Centre of foot pressure (CoP) displacements were recorded using the force platform for a period of 10 s in two experimental conditions of No pain and Pain. The No pain condition served as a control condition. In the Pain condition, painful electrical stimulations (stimulator Myodystim®) were applied to the muscle belly of the pars descendents of *Trapezius* muscles of both sides. To induce pain without causing any concomitant increased activation of neck muscles, a direct constant-current of 10 mA intensity was delivered via bipolar surface Ag-AgCl electrodes. The anode's side was determined at random. The electrical stimulation started 5 s prior to the 10 s postural data collection and ended 5 s after it to avoid the contamination of the variables computed from the CoP displacements by a possible startle effect related to the apparition and/or the cessation of the painful electrical stimulation (e.g. Corbeil et al., 2004).

A series of four consecutive trials were performed of each experimental condition of No pain and Pain. The order of presentation of these conditions was counterbalanced, with half of the subjects starting with the No pain condition and the other half starting with the Pain condition. Rest periods of 30 s were provided between each trial; 5 minutes period were between conditions.

At the end of each trial, subjects were asked to rate their perception of pain intensity to the electrical stimulation by drawing a line on a 10 cm visual analogue scale where 0 cm indicated "no pain" and 10 cm "intolerable pain". Visual analogue scales have been validated for experimental pain (Price et al., 1983).

**Signal processing**

The ground reaction forces, issued from three vertical mono-axial dynamometric load cells (range: 0–100 daN), were simultaneously recorded with a 64 Hz frequency on a personal computer during the tests (without any filtering). The signals were then amplified and converted from analogue to digital form through a 12 bits acquisition card. The CoP displacements along medio-lateral (ML) and antero-posterior (AP) axes were calculated from ground reaction forces.

The determination of the CoM displacements was derived from the CoP trajectories on a frequency basis through an amplitude ratio between CoM and CoP developed by Brenière (1996) for gait and adapted by Caron et al. (1997) for standing posture. The following steps detail the procedure to estimate the CoM:

- The CoP time series were transformed into the frequency domain by a discrete Fast Fourier Transform.

- The complex spectrum of the CoP time series was multiplied by the low-pass filter, CoM/CoP; which is defined as follows:

$$\text{CoM/CoP} = \Omega_0^2 / (\Omega_0^2 + \Omega^2),$$

where $\Omega = 2\pi f$ is the pulsation (rad.s$^{-1}$), and

$\Omega_0 = [mgh/(I_G + mh^2)]^{1/2}$ (Hz), termed natural body frequency, is a biomechanical constant relative to the anthropometry of the subject (m, g, h, $I_G$: mass of the subject, gravity acceleration, distance from the CoM to the ground, and the moment of body inertia around ML or anteroposterior AP axis with respect to the CoM).

- Depending on the direction, two distinct relationships were used to characterise the subject's anthropometry since the moments of inertia are slightly different. According to

Ledebt and Brenière (1994), these moments of inertia can be deduced from the following relationships:

$$I_G ML = 0.0572 m H_s^2 \quad \text{and} \quad I_G AP = 0.0533 m H_s^2,$$

where $H_s$ represents the height of the subjects.

- The filtered spectrum of the CoP was, thereafter, considered to be equal to the spectrum of the CoM time series. An inverse discrete Fast Fourier Transform was used to obtain the trajectory of the CoM in the time domain.

Data analysis

CoP and CoM displacements were processed along the ML and AP axes through two different analyses:

(1) a space-time-domain analysis first included the calculation of the variance (mm²), range (mm) and mean velocity (mm/s);

(2) a frequency-domain analysis, issued from the Fast Fourier Transform process, included the calculation of the mean and median frequencies (Hz) characterising the frequency components of the sway movements, the mean frequency representing the centroid of the spectrum and the median frequency separating the power spectrum into two equal energy areas.

Statistical analysis

We first controlled whether the order of testing has an effect on the pain intensity and postural assessments. Dependant variables were submitted to separate one-way analyses of variance (ANOVAs) (2 Orders of testing (No pain / Pain, Pain / No pain)). Level of significance was set at 0.05. Results of these preliminary analyses showed no significant main effect of order of testing for each dependent variable (*Ps*>0.05). Data from the two groups of subjects were then collapsed for further analyses.

Dependant variables were then submitted to separate two-way ANOVAs (2 Pain (No pain *vs.* Pain) × 2 Axes (Medio-lateral *vs.* Antero-posterior) × 4 trials (Trial 1 *vs.* Trial 2 *vs.* Trial 3 *vs.* Trial 4) with repeated measures on all factors.

**Results**

Perceived pain intensity

All subjects reported no pain for the *No pain* condition.

For the *Pain* condition, the mean visual analogue scores of the electrical stimulation was 7.1 ± 1.7 cm. The visual analogue scores were not different for all trials ($P>0.05$). All subjects described the electrical stimulation as a distinct pricking pain confined to a small area around the site of stimulation (i.e., pars descendents of *Trapezius* muscles of both sides) without spread.

Postural analyses

CoP displacements

Space-time-domain analysis

-------------------------------------

Please insert Figure 1 about here

-------------------------------------

Analysis of the variance of the CoP displacements showed main effects of Axis ($F(1,15) = 16.59$, $P<0.01$) and Pain ($F(1,15) = 6.25$, $P<0.05$), yielding larger variances along the AP than ML axis, and in the Pain than No pain condition (Figure 1A), respectively.

Analysis of the range of the CoP displacements showed main effects of Axis ($F(1,15) = 25.64$, $P<0.001$) and Pain ($F(1,15) = 14.93$, $P<0.01$), yielding larger ranges along the AP than ML axis, and in the Pain than No pain condition (Figure 1B), respectively.

Analysis of the mean velocity of the CoP displacements showed main effects of Axis ($F(1,15) = 42.84$, $P<0.001$) of Pain ($F(1,15) = 22.12$, $P<0.001$), yielding faster mean velocities along the AP than ML axis, and in the Pain than No pain condition (Figure 1C), respectively.

### Frequency-domain analysis

Analysis of the mean frequency of the CoP displacements showed a main effect of Pain ($F(1,15) = 14.10$, $P<0.01$), yielding higher mean frequency in the Pain than No pain condition (Figure 1D).

Analysis of the median frequency of the CoP displacements showed a main effect of Pain ($F(1,15) = 15.21$, $P<0.01$), yielding higher median frequency in the Pain than No pain condition (Figure 1E).

## CoM displacements

### Space-time-domain analysis

-------------------------------------

Please insert Figure 2 about here

-------------------------------------

Analysis of the variance of the CoM displacements showed main effects of Axis ($F(1,15) = 16.53$, $P<0.01$) and Pain ($F(1,15) = 8.04$, $P<0.05$), yielding larger variances along the AP than ML axis, and in the Pain than No pain condition (Figure 2A), respectively.

Analysis of the range of the CoM displacements showed main effects of Axis ($F(1,15) = 28.68$, $P<0.001$) and Pain ($F(1,15) = 14.97$, $P<0.01$), yielding larger ranges along the AP than ML axis, and in the Pain than No pain condition (Figure 2B), respectively.

Analysis of the mean velocity of the CoM displacements showed main effects of Axis ($F(1,15) = 65.81$, $P<0.001$)of Pain ($F(1,15) = 21.14$, $P<0.001$), yielding faster mean velocities along the AP than ML axis, and in the Pain than No pain condition (Figure 2C), respectively.

Frequency-domain analysis

Analysis of the mean frequency of the CoP displacements showed main effects of Axis ($F(1,15) = 10.55$, $P<0.01$) and Pain ($F(1,15) = 9.49$, $P<0.01$), yielding higher mean frequencies along the AP than ML axis, and in the Pain than No pain condition (Figure 2D), respectively.

Analysis of the median frequency of the CoP displacements showed main effects of Axis ($F(1,15) = 16.98$, $P<0.001$) and Pain ($F(1,15) = 9.38$, $P<0.01$), yielding higher median frequencies along the AP than ML axis, and in the Pain than No pain condition (Figure 2E)., respectively.

**Discussion**

The purpose of the present experiment was to investigate the effect of experimentally induced neck muscle pain via electrical stimulation on standing balance in humans. Postural control and postural performance were assessed by centre of foot pressure (CoP) displacements and centre of mass (CoM) displacements, respectively (e.g., Winter, 1995).

Before considering the effects on postural behaviour, it is important to mention that the visual analogue scores clearly indicated that the electrical stimulation used in the Pain condition of the present experiment was painful.

Analyses of the CoP and CoM displacements showed that experimental neck muscle pain deteriorated postural control and postural performance during quiet standing, respectively. In addition to the absence of significant interaction of Pain × Trial, Pain × Axis, and Pain × Axis × Trials, these results indicated that this destabilizing effect of experimental neck muscle pain was constant across all trials and similar along both the ML and AP axes.

The increased CoP and CoM displacements variances and ranges (Figure 1A,B and Figure 2A,B, respectively) observed in the Pain relative to the No pain condition suggested a less stable and more variable postural behaviour with than without neck muscle pain. These increased CoP and CoM displacements amplitudes induced by experimental neck muscle pain were accompanied by concomitant (1) faster CoP and CoM displacements, as indicated by increased mean velocities observed in the Pain relative to the No pain condition (Figure 1C and Figure 2C, respectively), and (2) increased CoP and CoM displacements frequencies, as indicated by the greater mean and median frequencies observed in the Pain than No pain condition (Figure 1D,F and Figure 2E,F, respectively).

On the whole, these results are in accordance with the decreased postural control observed in patients with chronic neck pain compared with matched free-pain control subjects (e.g., Alund et al., 1993; Karlberg et al., 1995; Madeleine et al., 2004; Michaelson et al., 2003; Persson et al., 1996; Poole et al., 2008; Treleaven et al., 2005). However, as mentioned above (Bennell and Hinman, 2005; Bennell et al., 2005), causality cannot be determined in clinical cross-sectional studies since any confounding effect of pain with a subjacent pathology and adaptive adjustments of motor control strategies (e.g., Latash and Anson, 1996. Vuillerme et al., 2001) can not be *a priori* excluded. Conversely, using an experimental pain model (Madeleine et al., 1999) to provide painful stimulation of the neck muscles in healthy individuals allows the effects of pain *per se* on postural control to be assessed (Bennell and Hinman, 2005; Bennell et al., 2005; Madeleine et al., 1999).

Interestingly, by showing increased CoP and CoM displacements in the Pain relative to the No pain condition (Figures 1 and 2), the present results are not in line with a recent investigation reporting that experimental neck muscle pain did not perturb postural control during quiet standing in healthy adults (Madeleine et al., 2004). The authors, however, suggested that the pain was probably too limited in intensity and spreading to decrease postural control. In Madeleine et al. experiment, experimental pain was induced unilaterally by a single bolus injection of hypertonic saline onto the right *Trapezius* muscle 30 s prior to the first postural trial and yielded a "moderate" perceived intensity of pain (mean visual analogue score of $4.3 \pm 0.4$ cm). In our experiment, electrical painful stimulations were applied bilaterally to the *Trapezius* muscles of both sides, lasted the entire duration of each postural trial and yielded a higher ("strong") perceived intensity of pain (mean visual analogue scores of $7.1 \pm 1.7$ cm). It is thus possible that, contrary to Madeleine et al. experiment, our painful stimulation was sufficient enough to have an effect on postural control during quiet standing. Interestingly, although the locus of painful stimulation was different, this suggestion is supported by two recent studies evidencing that painful stimulation of relatively high intensity is necessary to deteriorate upright postural control (Blouin et al., 2003; Corbeil et al., 2004). Indeed, a weak intensity of the painful stimulation either applied to the calves (Blouin et al., 2003) or feet (Corbeil et al., 2004) did not affect CoP displacements, whereas increasing the intensity of the painful stimulation gradually increased CoP displacements.

So, what could be the possible reasons leading to this observed destabilizing effect of experimental neck muscle pain? At least, three hypotheses can be formulated to account for the present set of data.

Assuming that pain interrupts, distracts and demands attention (e.g., Crombez et al., 1996; Eccleston and Crombez, 1999), a first hypothesis suggests that the painful stimulation may have influenced upright postural control via attentional effects. At this point, however, two separate set of data lead us to exclude this hypothesis. On the one hand, numerous studies have reported decreased CoP displacements when standing young healthy adults were required to perform a concurrent cognitive task relative to when they were asked to focus on their postural control (e.g., Ehrenfried et al., 2003; Hunter and Hoffman, 2001; Riley et al., 2003; Swan et al., 2007; Vuillerme et al., 2000; Vuillerme and Vincent, 2006), whereas the present results showed increased CoP displacements in the Pain relative to the No pain condition (Figures 1 and 2). On the other hand, results of a recent experiment showing that, for a similar perceived pain, painful stimulation applied to the hands did not increase the CoP displacements, whereas painful stimulation applied to the feet did (Corbeil et al., 2004), suggested that a sensory perception of a noxious stimulation *per se*, once it had captured attention (e.g., Crombez et al., 1996; Eccleston and Crombez, 1999), did not modify postural control during quiet standing.

According to a second hypothesis, the postural effects of experimental neck muscle pain may have been mediated through changes in arousal and/or anxiety, which previously has been shown to modify the way in which balance is controlled during quiet standing (e.g., Brown et al., 2006; Carpenter et al., 1999, 2001; Maki and McIlroy 1996). Based on the inverted pendulum model to represent quiet standing postural control (e.g., Gage et al., 2004), it was hypothesized that the combined observation of decreased amplitude and increased mean power frequency of the CoP displacements represents the adoption of an ankle stiffening strategy in response to increased perceived postural threat. The observation of increased variance (Figure 1A) and range of the CoP displacements (Figure 1B) in the Pain relative to the No pain condition suggests that the application of painful stimulation to the neck muscles did not result in a tighter control of posture and the adoption of a stiffening strategy.

Finally, according to a third hypothesis, the painful stimulation may have disturbed postural control during quiet standing through sensorimotor mechanisms (Blouin et al., 2003; Corbeil et al., 2004). At this point, it is reasonable to hypothesize that altered proprioceptive feedback from the neck muscles could contribute to the postural impairment observed in the Pain relative to the No pain condition (Brandt and Bronstein, 2001; Michaelson et al., 2003; Persson et al., 1996). Although the locus of the painful stimulation was different, this suggestion is supported by two recent studies. Results of the first study showed that muscle pain could distort proprioception at the interphalangeal joint of the thumb (Weerakkody et al., 2008). Interestingly, these results further evidenced that that this disturbing effect depended on the site at which the pain was initiated, i.e., involved regions and tissues that have a proprioceptive role at the joint. Results of the second study reported impaired ankle joint proprioceptive acuity when painful stimulations were applied to ankle muscles (Matre et al., 2002). These authors suggested that experimental muscle pain could have decreased the sensitivity in muscle spindle receptors, and/or altered thresholds of spinal interneurones to sensory signals and/or reduced the ability to interpret proprioceptive signals. However, without direct measures of the neck proprioception in asymptomatic subjects, such a proposal is yet speculative and warrants additional investigations. More largely, with regard to the hypothesis of an impairment of neck neuromuscular function induced by the painful stimulation, our results are in line with previous studies which also have reported degraded postural control during quiet standing when neck proprioception was specifically altered through experimental manipulations in asymptomatic subjects, such as muscular vibration (e.g., Gregoric et al., 1978; Koskimies et al., 1997; Roll and Roll, 1988) or muscle fatigue (Gosselin et al., 2004; Schieppati et al., 2003; Vuillerme et al., 2005).

In conclusion, the results of the present experiment showed that experimental neck muscle pain significantly impaired standing balance in humans. These findings thus emphasize the destabilizing effect of neck muscle pain *per se* and, more largely, confirm the importance of intact neck neuromuscular function on the maintenance of an erect stance. Further research is now needed to investigate the generalization of these results to different stance conditions, sensory environments, cognitive contexts and postural tasks.

**Acknowledgements**

The authors would like to thank subject volunteers and the anonymous reviewers for their valuable comments and suggestions on the first version of the manuscript.

**Figure captions**

**Figure 1.** Mean and standard error of mean of the variance (A), range (B), mean velocity (C), mean frequency (D) and median frequency (E) of the centre of foot pressure (CoP) displacements obtained in the two conditions of No pain and Pain of neck muscles. The two experimental conditions are presented with different symbols: No pain (*black bars*) and Pain (*white bars*). The significant *P*-values for comparisons between No pain and Pain conditions also are reported (*: $P<0.05$; **: $P<0.01$; ***: $P<0.001$).

**Figure 2.** Mean and standard error of mean of the variance (A), range (B), mean velocity (C), mean frequency (D) and median frequency (E) of the centre of mass (CoM) displacements obtained in the two conditions of No pain and Pain of neck muscles. The two experimental conditions are presented with different symbols: No pain (*black bars*) and Pain (*white bars*). The significant *P*-values for comparisons between No pain and Pain conditions also are reported (*: $P<0.05$; **: $P<0.01$; ***: $P<0.001$).

**Figure 1**

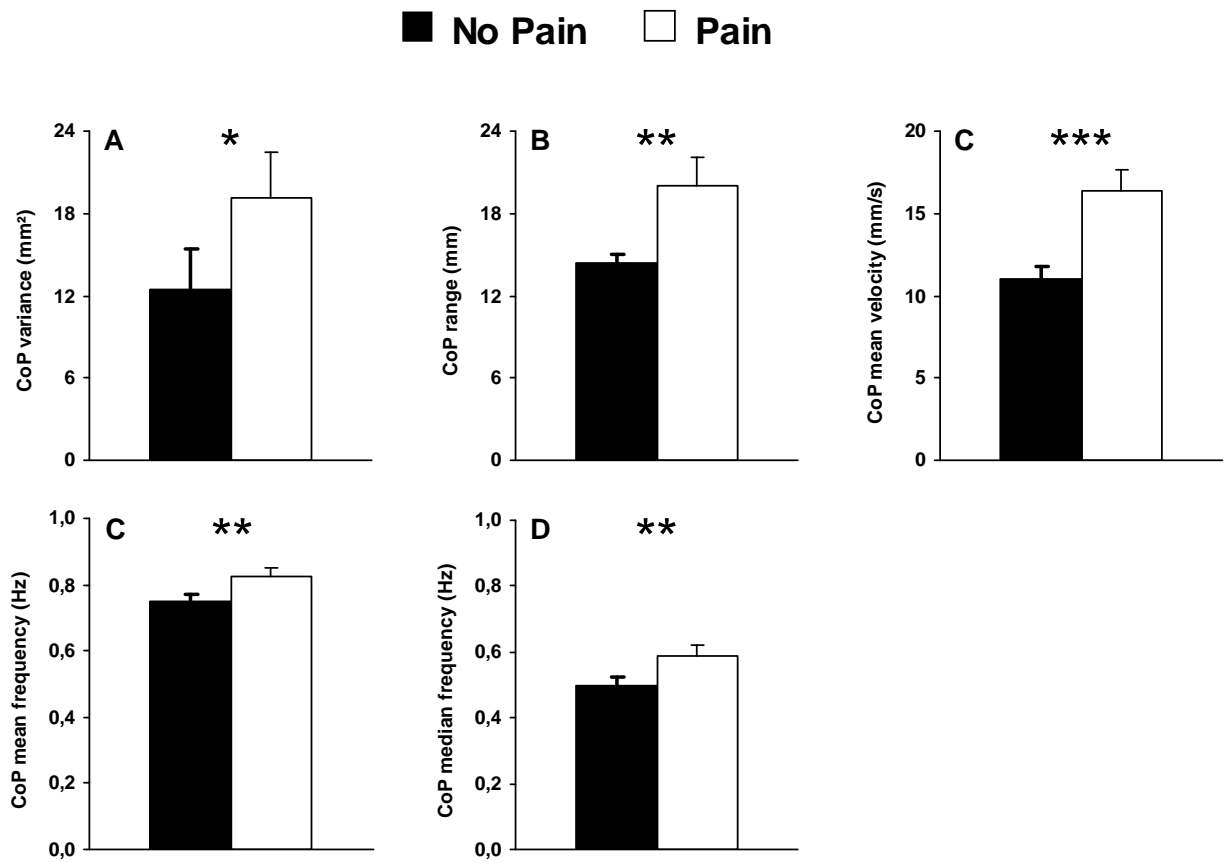

**Figure 2**

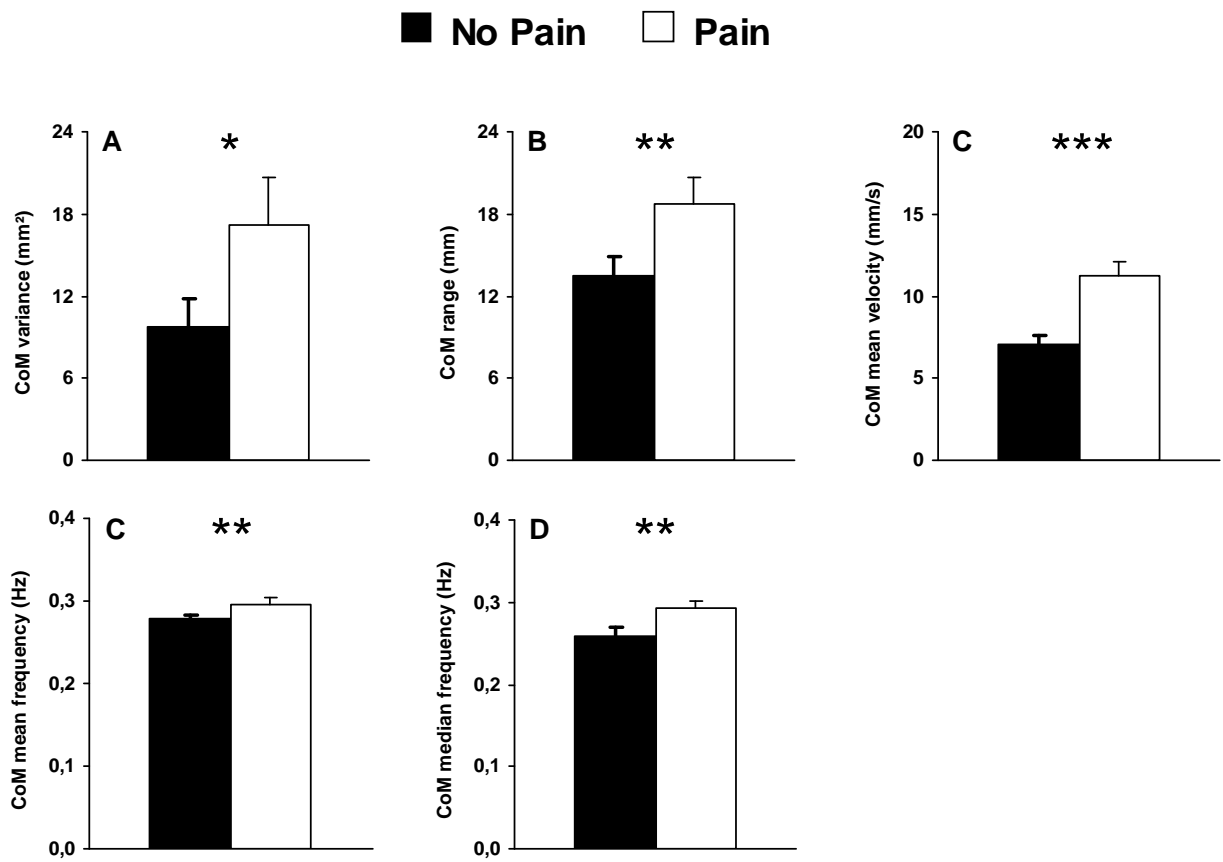